\documentclass [12pt]{article}
\usepackage{amsmath,amssymb,epsfig}
\usepackage{graphicx}

\RequirePackage[english, russian]{babel}
\RequirePackage[cp1251]{inputenc}

\begin{document}
\def\refname{\Large~~~~~~~{\bf References}}
\newcommand{\el}{\left}
\newcommand{\er}{\right}
\newcommand{\p}{\prime}
\newcommand{\rr}{\rho}
\newcommand{\ro}{\rho^\circ}
\newcommand{\ti}{\tilde}
\newcommand{\veps}{\varepsilon}
\newcommand{\dis}{\displaystyle}
\newcommand{\scr}{\scriptsize}
\begin{center}
{\Large {\bf  EXCITATION OF NUCLEAR COLLECTIVE STATES \\ BY HEAVY
IONS WITHIN THE MODEL OF \\ [2mm] SEMI-MICROSCOPIC  OPTICAL
POTENTIAL
}} \\[5mm]
{\large\bf  ~K.M.~Hanna$^{1}$, K.V~Lukyanov$^2$,
V.K.~Lukyanov$^2$, Z.Metawei$^{3}$ B.~S{\l}owi{\'n}ski$^{4,5}$,
E.V.~Zemlyanaya$^2$}\\[3mm]
{\small\it ~~~~~$^1$Math. and Theor. Phys. Dept., NRC, Atomic Energy
Authority, Cairo, Egypt}\\
{\small\it $^2$Joint Institute for Nuclear Research, Dubna,
Russia~~~~~~~~~~~~~~~~~~~~~~~~~~~~}\\
{\small\it $^3$Physics Department, Faculty of Science, Cairo University,
        Giza, Egypt~~~~}\\
{\small\it $^4$Faculty of Physics, Warsaw University of
Technology, Warsaw, Poland ~~~}\\
{\small\it $^5$Institute of Atomic Energy, Otwock-Swierk,
Poland~~~~~~~~~~~~~~~~~~~~~~~~~~~~}\\
\vspace{0.5cm}
\end{center}
\vspace{0.5cm}

{\small
The (semi-)microscopic double-folding nucleus-nucleus optical potentials are
suggested for consideration of inelastic scattering with excitation of
collective nuclear states by using the adiabatic approach and the elastic
scattering amplitude in the high-energy approximation. The analytical
expression for inelastic scattering amplitude is obtained keeping
the first order terms in the deformation parameter of a potential.
Calculations of inelastic cross sections for the $^{17}$O heavy ions scattered
on different nuclei at about hundred Mev/nucleon are made, and the acceptable
qualitative agreement with the experimental data is obtained without
introducing free parameters. The prospect of the method for further
applications is discussed.
}

\vspace{0.5cm}

\def\baselinestrech{1.5}

{\normalsize
\section {Introduction}

The theory of excitations of nuclear collective states in peripheral
nuclear collisions is based on the elastic scattering optical potential
$U(r)=V(r)+iW(r)$. This latter is used to obtain the transition potential
$U_{int}=U_{int}^{(N)}+U_{int}^{(C)}$ for inelastic channel. Recently, in
\cite{LMZ}, the nucleus-nucleus inelastic scattering with excitation of
2$^+$ rotational states was considered in the framework of the high-energy
approximation utilizing the phenomenological Woods-Saxon type potential.
The collective variables $\{\alpha_{\lambda\,\mu}\}$, which characterize the
deformation of the surface of a potential, were introduced through the radius
\begin{equation}\label{p2_1}
\Re\, =\,R\,+\,\delta R, \qquad
\delta R\,=\,R\sum\limits_{\lambda\mu}\,\alpha_{\lambda\mu}\,
Y_{\lambda\mu}(\theta,\phi).
\end{equation}
Here $\theta,\phi$ are spherical coordinates of a space vector ${\bf r}$
in the laboratory system. The wave functions of rotational states and
collective variables $\{\alpha_{\lambda \,\mu}\}$ are given as follows:
\begin{equation}\label{p2_2}
|I\,M>=\sqrt{2I+1\over 8\pi^2}\,D^{(I)}_{M0}(\Theta_i), \qquad
\alpha_{2\,\mu}\,=\,\beta_2 \,D^{(2)\,*}_{\mu 0}(\Theta_i),
\end{equation}
where $\beta_2$ is the deformation parameter and $\{\Theta_i\}$ are
the intrinsic axis rotational angles.

In \cite{LMZ}, suggesting small $\beta_2\ll 1$, the transition potential was
obtained as the derivative of $U(r,\Re)$, and the inelastic scattering
amplitude was derived in adiabatic approximation
\begin{equation}\label{p2_3}
f_{IM}(q)\, = \,<IM|\,f(q,\{\alpha_{\lambda\mu}\})|00>,
\end{equation}
where $q=2k\sin(\vartheta/2)$ is the momentum transfer, $k$ is the relative
momentum, and $\vartheta$, the angle of scattering. The elastic scattering
amplitude $f(q,\{\alpha_{\lambda\mu}\})$ was taken in the high-energy
approximation (HEA) with the "frozen" coordinates of collective motion
$\{\alpha_{\lambda\mu}\}$. Then, inelastic cross sections for the $^{17}$O
heavy ions scattered on different nuclei at about hundred Mev/nucleon were
calculated, and an acceptable agreement with the experimental data was
received. So, the conclusions were made on applicability of HEA to study
the nucleus-nucleus inelastic processes.

The aim of this paper is to apply not phenomenological but microscopic
potentials for calculating an inelastic scattering amplitude. The matter
of fact is that the phenomenological potentials, used for inelastic
scattering, must be specially fitted in the corresponding elastic channel
at the same energy and for the same couple of scattered nuclei as they are
in inelastic channel. Otherwise, at present there are no tables of global
optical potentials for the heavy-ion elastic scattering at different energies
and kinds of colliding nuclei. Moreover, there exists the problem of ambiguity
of parameters of phenomenological potentials (see, e.g., \cite{Sat}) since
the fit needs a large amount of data, and thus any additional
information, involved into consideration, in particular, the data
of inelastic scattering, is very desirable.

On the other hand, in the last two decades of years the double-folding (DF)
microscopic nucleus-nucleus potentials occur rather popular.
They are calculated using the following expression:
$$
V^{DF}(r)\,=\,V^D(E,r)\,+\,V^{EX}(E,r)\,=\,\int d^3 r_1 \, d^3 r_2 \,
\rho_1({\vec r}_1)\, \rho_2({\vec r}_2)\, v^D(\rho, E, r_{12})\,+
$$
\begin{equation}\label{p2_4}
+\,\int d^3 r_1 \, d^3 r_2 \,
\rho_1({\vec r}_1, {\vec r}_1+{\vec r}_{12})\,
\rho_2({\vec r}_2, {\vec r}_2-{\vec r}_{12})
v^{EX}(\rho, E, r_{12})\,\exp\el[{i{\vec k}(\vec r){\vec r}_{12}\over M}\er],
\end{equation}
where ${\vec r}_{12}={\vec r} + {\vec r}_2 - {\vec r}_1$ is the distance
between colliding nuclei, $k(r)$ is the local momentum of relative motion
of nuclei, and $M=A_1A_2/(A_1+A_2)$. (For details see, e.g., \cite{KK90},
\cite{KS00}). These DF-potentials apply the nuclear density distributions
$\rho(\vec r)$ and matrices $\rho(\vec x,\vec y)$, and also include the
effective nucleon-nucleon potentials $v^D$ and $v^{EX}$. In principle, all of
these quantities are known from independent experimental studies. It was
also established dependence of NN-potentials on kinetic energy and on the
matter density in overlapping region of nuclei. DF-potentials take into
account effects of exchange and break up of nucleons, and describe
sufficiently well the shape of the peripheral region of potentials, very
important in formations both the differential and total cross-sections.
For a period of years, in comparisons with experimental data, these
real DF-potentials were supplemented by the phenomenological imaginary
potentials $W^P(r)$ having three (or more) free parameters. By doing so,
it was shown that one needs to diminish slightly the calculated real part by
introducing a renormalization coefficient $N_r$, and, thus the whole
potential $U(r)=N_rV^{DF}+iW^P(r)$ has four (or more) free parameters.

However, recently in \cite{LZL}, it was demonstrated that the imaginary
part can be also calculated microscopically by transforming the eikonal
phase of the high-energy microscopical theory \cite{{Gla},{Sit}} of
scattering of complex systems. It was shown in \cite{LZL} that this
imaginary potential $W^H$ contains the folding integral which corresponds to
the integral of only the direct part $V^D$ of the DF-potential (\ref{p2_4}).
This optical potential has the form $U(r)=N_RV^{DF}(r)+iN_{im}W^H(r)$.
In addition, it was reasonable to generalize this form to include the
exchange term, too, and then to test the potential $U(r)=N_rV^{DF}+
iN_{im}V^{DF}(r)$, as well. These potentials were called the semi-microscopic
ones because of their basic forms $W^H$ and $V^{DF}$ were calculated
microscopically, without introducing free parameters, and only two parameters
$N_r$ and $N_{im}$ must be adjusted to experimental data.

\begin{figure}
\begin{center}
\includegraphics[ width = 0.8\linewidth]{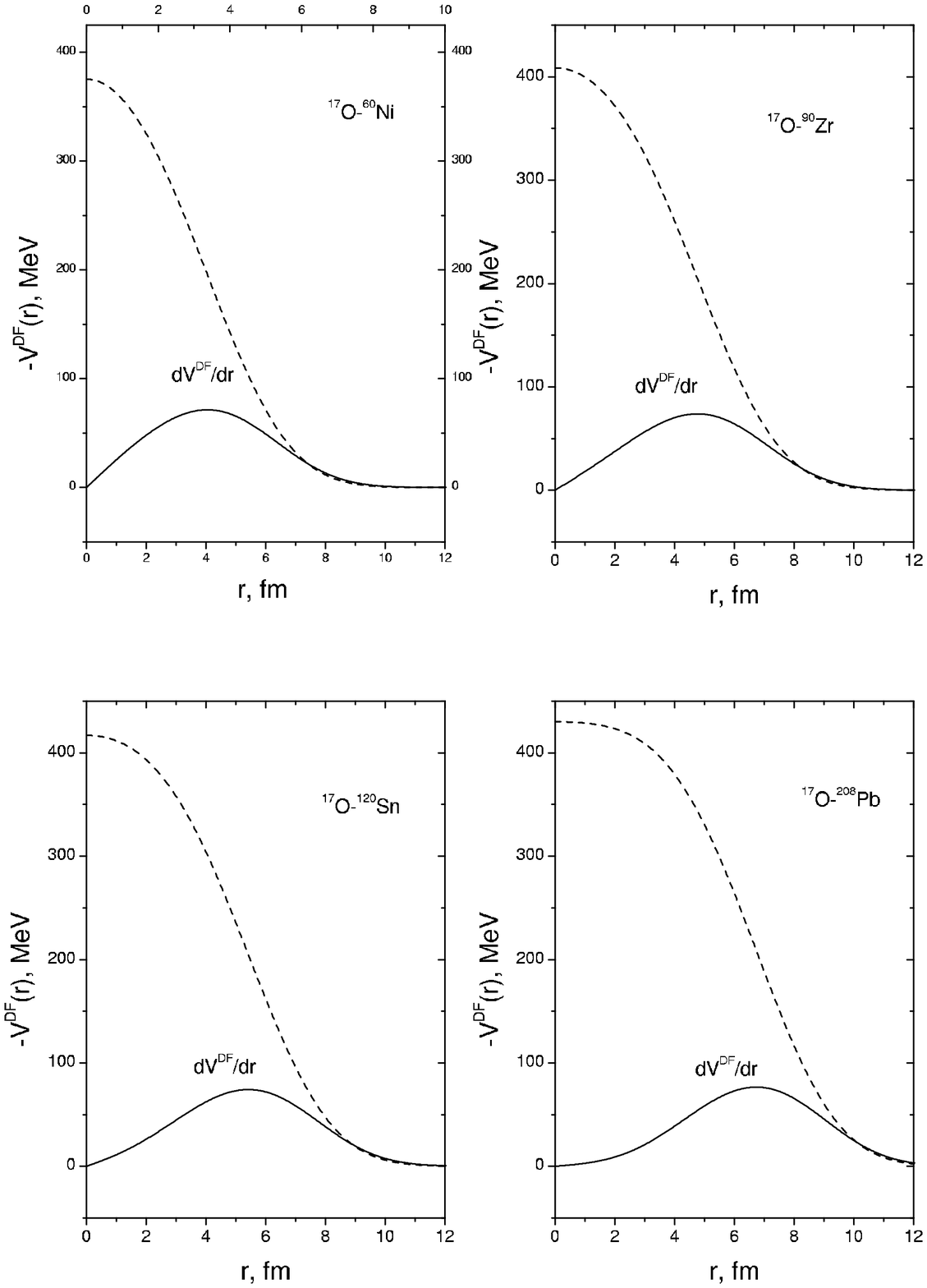}
\end{center}
 Fig. 1. {The double-folding potentials (dashed
lines) and their derivatives (solid lines) calculated for
different couples of nuclei at $E_{lab}$=1435 MeV. }
\end{figure}

\begin{figure}
\begin{center}
\includegraphics[ width = 0.8\linewidth]{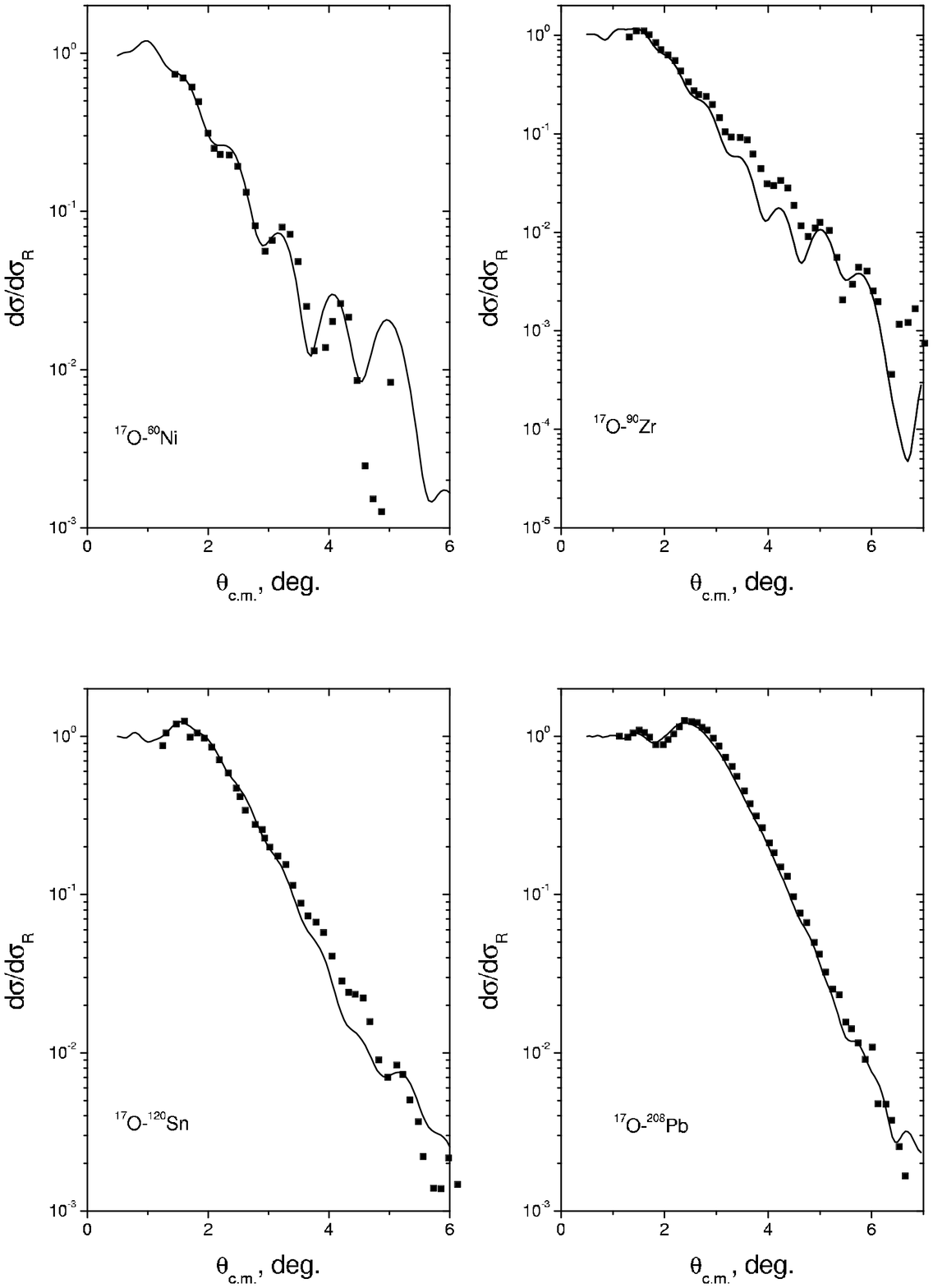}
\end{center}
Fig. 2. {The ratio of the elastic scattering differential cross
sections to the Rutherford one (solid lines) calculated using the
semi-microscopic optical potentials $N_rV^{DF}+iN_{im}V^{DF}$ at
$E_{lab}$=1435 MeV and compared with the experimental data from \cite{Neto}.
 }
\end{figure}

\begin{figure}
\begin{center}
\includegraphics[ width = 0.8\linewidth]{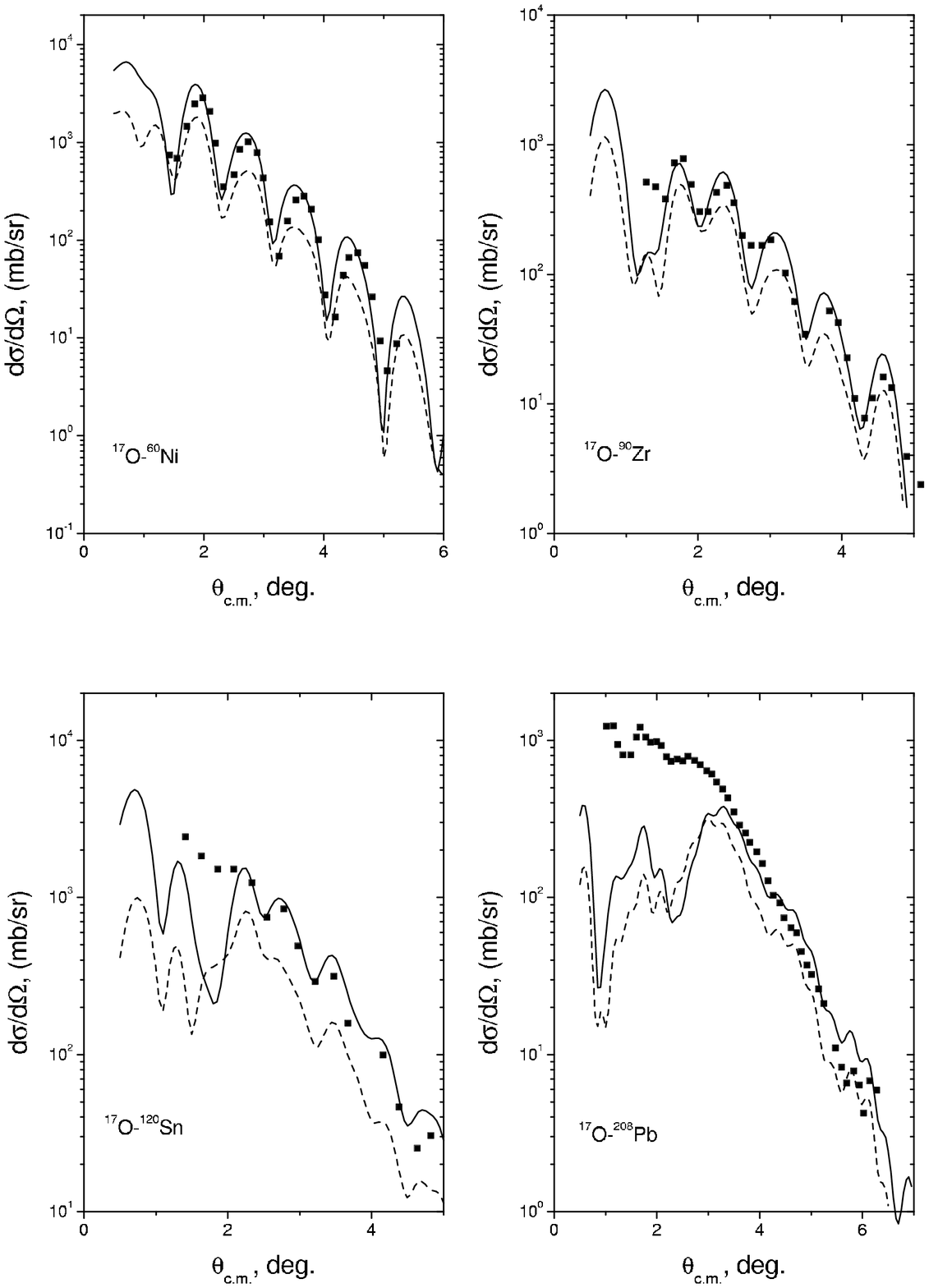}
\end{center}
Fig. 3. {Microscopic calculations of inelastic scattering
differential cross-sections of the ${^{17}O}$ heavy ions at 1435
MeV on different target nuclei with excitations of 2$^+$
collective states. Dashed curves are for the consistent
deformation parameters of the nuclear and Coulomb potentials (set
1 in Table), solid curves - with the free $\beta^{(n)}_2$
parameters (set 2). The data are from \cite{Neto}.
   }
\end{figure}

Figure 1 shows by dashed lines the double-folding potentials $V^{DF}$
calculated in \cite{LZLea} for scattering of the heavy ions $^{17}$O on
different nuclei at E$_{lab}$=1435 MeV. The respective optical potentials
were adjusted to the elastic scattering differential cross sections using
for imaginary terms the forms $N_{im}W^H$ and $N_{im}V^{DF}$. In Fig.2 we
reproduce the ratios of elastic cross sections to the Rutherford one
calculated in \cite{LZLea} in the framework of the high-energy
approximation using the microscopic potentials $U(r)=N_rV^{DF}(r)+
iN_{im}V^{DF}$, and their comparisons with the experimental data from
\cite{Neto}. The adjusted normalization coefficients $N_r$ and $N_{im}$ were
obtained as 0.6 and 0.6 for $^{60}$Ni, 0.6 and 0.5 for $^{90}$Zr, 0.5 and 0.5
for $^{120}$Sr, and 0.5 and 0.8 for $^{208}$Pb. One sees fairly well agreement
with the data in the region of an applicability of HEA at $\theta\leq
\sqrt{2/kR}$. So, these potentials can be applied further in calculations
of inelastic scattering of the same nuclei at the same energy for comparisons
with existent experimental data on the 2$^+$ state excitations \cite{Neto}.

\section {Some formulae and comments}

The microscopic potentials have no the obvious parameters something like
the radius $R$ and diffuseness $a$ of a Woods-Saxon potential. Therefore,
in order to introduce there the dependence on internal collective variables
$\alpha_{\lambda\,\mu}$, we make, in analogy with (\ref{p2_1}), the
respective changes of spatial coordinates
\begin{equation}\label{p2_5}
r\,\Rightarrow \, r\,+\,\delta r, \qquad \delta
r\,=\,-\,r\sum\limits_{\lambda\mu}\,\alpha_{\lambda\mu}\,
Y_{\lambda\mu}(\theta,\phi).
\end{equation}
 Then, expanding the potential in $\delta r$ we obtain the
generalized optical potential consisted of two terms, the
spherically symmetrical and deformed one
\begin{equation}\label{p2_6}
U^{(N)}(r, \{\alpha_{\lambda\,\mu}\})\,=\,U^{(N)}(r)\,+\,
U^{(N)}_{int}(r, \{\alpha_{\lambda\,\mu}\}),
\end{equation}
 where the transition potential (its nuclear part) is as follows
\begin{equation}\label{p2_7}
U^{(N)}_{int}\,=\,-\,r\,{d\over dr}\,U(r)
\sum\limits_\mu\,\alpha_{2\mu} \, Y_{2\mu}(\theta,\phi).
\end{equation}
 In Fig.1 one can see behavior of the derivatives of microscopic
potentials $V^{DF}$ for the above considered cases. All of them
have a typical maxima in the surface region of a potential. The
respective quadrupole part of the generalized Coulomb potential
$U^{(C)}(r, \{\alpha_{\lambda\,\mu}\})$ is obtained as usually
with a help of its definition through the uniform charge density
distribution having the radius $\Re$ as in (\ref{p2_1}) with
$R_C=r_c(A_1^{1/3}+A_2^{1/3})$. This yields at \cite{LMZ}
\begin{equation}\label{p2_8}
U^{(C)}_{int}\,=\,{3\over 5}\,U_B\,
\Bigl[\Bigl({r\over R_C}\Bigr)^2\Theta(R-r)+\Bigl({R_C\over r}\Bigr)^3
\Theta(r-R)\Bigr]\,\sum\limits_\mu\, \alpha_{2\mu}
\,Y_{2\mu}(\theta,\phi),
\end{equation}
where ~$U_B=Z_1Z_2e^2/R_C$.

Then, we use the expression for high-energy amplitude of scattering
\begin{equation}\label{p2_9}
f(q)\, =\, i{k\over 2\pi} \int \, bdbd\phi \, {\dis e}^{\dis iqb\cos\phi}
\, \Bigl [1-{\dis e}^{\dis i\Phi}\Bigr ].
\end{equation}
Here integration is performed over impact parameters $b$ and on
its azimuthal angle $\phi$, and the eikonal phase is determined by
the nucleus-nucleus potential
\begin{equation}\label{p2_10}
\Phi\,=\,-{1\over \hbar v}\,\int_{-\infty}^\infty U(r+\delta r)\,dz,
\qquad r=\sqrt{b^2+z^2},
\end{equation}
 where $v$ is the relative velocity of colliding nuclei.
Substituting here the total potential having the central and
transition terms, one can write
\begin{equation}\label{p2_11}
\Phi\,=\,\Phi_0(b)\,+\,
\Phi_{int}(b,\,\{\alpha_{\lambda\mu}\},\,\phi).
\end{equation}
\begin{equation}\label{p2_12}
\Phi_{int}\,=\,\beta_2\,\sum\limits_{\mu=0,\pm 2}\,G_{\mu}(b)\,
D^{(2)\,*}_{\mu 0}(\Theta_i)\,{\dis e}^{\dis i\mu\phi}
\end{equation}
$$
 G_{\mu}(b)\,=\,-{2\over \hbar v}\,\int_0^\infty dz\,
Y_{2\mu}\el(\arccos(z/r),0\er)\times \hspace{7cm}
$$
\begin{equation}\label{p2_13}
\hspace{2cm}\times\el[\,-\,r{dU(r)\over dr}\,+\,{3\over 5}\,U_B
\Bigl[\Bigl({r\over R_C}\Bigr)^2\Theta(R-r)+\Bigl({R_C\over
r}\Bigr)^3 \Theta(r-R)\Bigr]\er]
\end{equation}
where $r=\sqrt{b^2+z^2}$. Substituting (\ref{p2_11}) in (\ref{p2_9}) and
(\ref{p2_3}), and expanding the exponential function $\exp(i\Phi_{int})$
we retain only a term of the first order in $\beta_2$. Then, integration
over rotational angles $\Theta_i$ can be performed, and one gets the inelastic
scattering amplitudes $f_{\lambda\,\mu}(q)$ and differential cross section
as follows \cite{LMZ}:
\begin{equation}\label{p2_14}
f_{2\,0}(q)\,= \,{k\over\sqrt{5}}\,\beta_2 \int_0^\infty \, bdb\, J_0(qb)\,
G_0(b)\,{\dis e}^{\dis i\Phi_0(b)},
\end{equation}
\begin{equation}\label{p2_15}
f_{2\,2}(q)\,= \,-\,{k\over\sqrt{5}}\,\beta_2 \int_0^\infty \, bdb\, J_2(qb)\,
G_2(b)\,{\dis e}^{\dis i\Phi_0(b)},
\end{equation}
\begin{equation}\label{p2_16}
{d\sigma_{in}\over d\Omega}\,=\,|f_{2\,0}|^2\,+\,2\,|f_{2\,2}(q)|^2.
\end{equation}

\section {Comparison with experimental data. Summary}

When calculating the elastic and inelastic scattering amplitudes one has
to take into account the Coulomb distortion of the straight-ahead trajectory
situated in expressions of the high-energy theory. This is made by exchanging,
in the nuclear part of the phases $\Phi_0(b)$ and $\Phi_{int}(b)$, the impact
parameter $b$ by
the distance of the turning point in the Coulomb field of the point
charge, i.e.,  $b\Rightarrow b_c=\bar a+\sqrt{b^2+{\bar a}^2}$, where
$\bar a=Z_1Z_2e^2/\hbar vk$ is a half of closest approach distance at $b$=0.

Firstly, we estimate inelastic cross sections of scattering of
$^{17}$O on different nuclei without introducing any free
parameters. To this end we apply semi-microscopic optical
potentials $U=N_rV^{DF}+iN_{im}V^{DF}$ calculated and adjusted in
\cite{LZLea} to the experimental data on elastic scattering of the
same nuclei \cite{Neto}. The deformation parameters
$\beta^{(n)}_2$ and $\beta^{(c)}_2$ for nuclear and Coulomb
potentials, separately, were suggested to obey the relation
$\beta^{(c)}_2\,{\bar R}_C=\beta^{(n)}_2\,{\bar R}_n$, where $\bar
R$ are ${\it rms}$ radii. Qualitatively, this relation supposes an
equality of areas of rings on the $r$-plane, where the main
transition takes place. The $\beta^{(c)}_2$ deformations are taken
as they were extracted in \cite{Neto} using the known reduced
electric transition probabilities $B(E2\uparrow)$ in the target
nuclei. (For parameters see set 1 in Table).

\vspace{.5cm}
{\large
{\samepage \hspace*{0.3cm} { {\bf Table.}~~{ \bf Deformation
parameters of the Coulomb\\ \hspace*{5cm}and nuclear potentials }}

\begin{center}
\begin{tabular}{|c|c|c|c|c|c|}
\hline \hline

         &  $\beta_2$    &${^{17}O}+{^{60}Ni}$  &${^{17}O}+{^{90}Zr}$
                       &${^{17}O}+{^{120}Sn}$ &${^{17}O}+{^{208}Pb}$ \\
\hline \hline
 Set 1  &$\beta^{(c)}_2$& 0.2067 & 0.091  & 0.1075 & 0.0544 \\
\cline{2-6}

         &$\beta^{(n)}_2$& 0.2453 & 0.1072 & 0.1270 & 0.0644  \\
\hline \hline
 Set 2   &$\beta^{(c)}_2$& 0.2067 & 0.091  & 0.1075 & 0.0544 \\
\cline{2-6}
        &$\beta^{(n)}_2$&  0.4   &  0.16  & 0.25   & 0.1     \\
\hline \hline
\end{tabular}
\end{center}
}
}
\vspace{.5cm}

Figure 3 shows these results by dashed lines. We see that calculations
performed without free parameters are in a qualitative agreement with the
experimental data. The slopes of all curves are in coincidence with the
behavior of the data. As to the absolute values of cross sections, they
can be slightly improved by increasing the deformation parameters.
An exception is seen at small angles (very peripheral collisions) for
heavy nuclei $^{120}$Sn, $^{208}$Pb (large charges), where the multi-step
Coulomb excitation must give large contribution while in our consideration
we take into account only the first power terms of deformations $\beta$ in
amplitudes.

Meanwhile, if we refuse to fulfill the above relation between $\beta^{(c)}_2$
and $\beta^{(n)}_2$, and suppose the deformation of nuclear potential
$\beta^{(n)}_2$ to be free parameter (solid curves) then agreements with the
data become fairly better as compared to the preceding calculations.
(For parameters see set 2 in Table).

Summarizing the obtained results we note, that, first, the outlook for
the further work in utilizing (semi)-microscopic potentials, both the
real and imaginary one, for parametrization and analysis of experimental
data is rather attractive. Their applications do not need to introduce
so many parameters as they included in the case of phenomenological potentials.

Second, our consideration was based on the simple adiabatic approximation
and utilizing only the linear terms, in $\beta_2$, of the scattering amplitude.
But these potentials can be also applied in the more proper coupled channel
method where, in principle, all powers of terms with $\beta_2$ are taken
into account. By the way, the prospects are also exist of improving the
suggested approach by constructing the microscopic transition potentials
rather than that obtained above from the microscopic potentials themselves.\\

\begin{center}{\Large ACKNOWLEDGMENTS}
\end{center}

The co-authors V.K.L. and B.S. are grateful to the Infeld-Bogoliubov Program.
E.V.Z. and K.V.L. thank the Russian Foundation Basic Research
(project 06-01-00228).

\end{document}